\documentclass[a4paper,11pt]{article}
\usepackage{pos}

\title{Reconstruction of stereoscopic CTA events using deep learning with CTLearn}
\ShortTitle{CTA deep learning analysis with CTLearn}

\author*[a]{T. Miener}
\author[a]{D. Nieto}
\author[b]{A. Brill}
\author[c]{S. Spencer}
\author[a]{J. L. Contreras}
\forColl{CTA Consortium}

\affiliation[a]{EMFTEL department and IPARCOS, Universidad Complutense de Madrid,\\
Madrid, Spain}

\affiliation[b]{Columbia University, Physics Department,\\
New York, USA}

\affiliation[c]{University of Oxford, Department of Physics,\\
Oxford, United Kingdom}

\emailAdd{tmiener@ucm.es}

\abstract{The Cherenkov Telescope Array (CTA), conceived as an array of tens of imaging atmospheric Cherenkov telescopes (IACTs), is an international project for a next-generation ground-based gamma-ray observatory, aiming to improve on the sensitivity of current-generation instruments a factor of five to ten and provide energy coverage from 20 GeV to more than 300 TeV. Arrays of IACTs probe the very-high-energy gamma-ray sky. Their working principle consists of the simultaneous observation of air showers initiated by the interaction of very-high-energy gamma rays and cosmic rays with the atmosphere. Cherenkov photons induced by a given shower are focused onto the camera plane of the telescopes in the array, producing a multi-stereoscopic record of the event. This image contains the longitudinal development of the air shower, together with its spatial, temporal, and calorimetric information. The properties of the originating very-high-energy particle (type, energy, and incoming direction) can be inferred from those images by reconstructing the full event using machine learning techniques. In this contribution, we present a purely deep-learning driven, full-event reconstruction of simulated, stereoscopic IACT events using CTLearn. CTLearn is a package that includes modules for loading and manipulating IACT data and for running deep learning models, using pixel-wise camera data as input.}

\FullConference{37$^{\rm{th}}$ International Cosmic Ray Conference (ICRC 2021)\\
		July 12th -- 23rd, 2021\\
		Online -- Berlin, Germany}

\begin{document}
\maketitle

\section{Introduction}
\label{sec:introduction}

The Cherenkov Telescope Array (CTA)~\cite{Acharyya:2019nwy} is the next-generation ground-based gamma-ray observatory, aiming to improve on the sensitivity of current-generation instruments by a factor of five to ten and provide an energy coverage from 20 GeV to more than 300 TeV. CTA will consist of two arrays of tens of imaging atmospheric Cherenkov telescopes (IACTs) to be built in the Northern Hemisphere (La Palma, Canary Island, Spain) and in the Southern Hemisphere (near Cerro Paranal, Chile). Arrays of IACTs observe simultaneously the Cherenkov light induced by the showers of particles produced when very-high-energy (VHE; above 20 GeV) gamma rays or charged cosmic rays enter the atmosphere. Those Cherenkov photons are collected by the optical systems and focused onto cameras, producing a stereoscopic record of the event. The IACT images contain the longitudinal development of the air shower, together with its spatial, temporal, and calorimetric information.

The gamma-ray and cosmic-ray initiated showers can be distinguished from their morphological differences, translated into their IACT stereoscopic images. This distinction, dubbed particle or event classification, is crucial for IACTs since cosmic-ray events are their main background. The original approach to classify IACT events from their images was to extract handcrafted features, like the commonly used Hillas parameters~\cite{1985ICRC....3..445H}, and perform parameter-wise selection over the multidimensional space of those parameters. As a result of the improvement in available computational resources and algorithms over the past few decades, this original approach evolved into more sophisticated strategies where supervised learning algorithms like Random Forests~\cite{2008NIMPA.588..424A} or Boosted Decision Trees~\cite{2009APh....31..383O,2011APh....34..858B,krause2017improved} are trained on those handcrafted features, substantially improving the performance of the particle classification and, consequently, the sensitivity of the instruments. In addition, IACT data analysis methods also need to infer further properties of the gamma-ray events, namely, the energy and the incoming direction of the originating particles. This so-called full-event reconstruction could also be performed with deep convolutional neural networks (DCNs), a particular class of deep learning algorithms, which are currently the most successful machine learning methods for computer vision, excelling at image classification and regression among other tasks~\cite{Goodfellow-et-al-2016}. Rather than crafting the features by hand, these types of algorithms are capable of learning the feature extraction by themselves (representation learning). Therefore, DCNs can access all the information contained in the images, not only those condensed in handcrafted features extracted from those images.

Previous works have demonstrated the potential application of these algorithms for IACT event reconstruction \cite{2017arXiv170905889N, veritas_cnn, 2018arXiv181000592M, 2019APh...105...44S, 2019-icrc-hex, 2021spencer}. DCN-based monoscopic telescope performance and the application of DCNs on observational data from the first Large-Sized Telescope (LST-1 prototype) of CTA North is discussed in these proceedings elsewhere \citep{Grespan2021, Vuillaume2021}. As a natural continuation of this line of work, this contribution focuses on full-event reconstruction of MC-simulated stereoscopic events.

\section{CTA analysis workflow with deep learning}
\label{sec:workflow}

The CTA analysis workflow consists of several software blocks. The MC simulations and later the observational data are reduced with the \texttt{stage1-tool} of \texttt{ctapipe}\footnote{\href{https://github.com/cta-observatory/ctapipe}{https://github.com/cta-observatory/ctapipe}}~\citep{Noethe2021, ctapipe}, a prototype low-level data processing pipeline for CTA, and the resulting calibrated images, as well as their image parameters, are stored in the official CTA Data Level 1 (DL1) format. Data loading and pre-processing, specially designed for deep learning purposes, are managed using an associated external package, \texttt{DL1-Data-Handler}\footnote{\href{https://github.com/cta-observatory/dl1-data-handler}{https://github.com/cta-observatory/dl1-data-handler}}~\citep{dl1dh}. It supports event-wise data reading using generators to handle big datasets. The training of the deep learning models and their inference, the actual full-event reconstruction, are performed with \texttt{CTLearn}\footnote{\href{https://github.com/ctlearn-project/ctlearn}{https://github.com/ctlearn-project/ctlearn}}~\citep{ctlearn, 2019-icrc-ctlearn}. The high-level products like instrument response functions (IRFs) and sensitivity curves are obtained using \texttt{pyirf}\footnote{\href{https://github.com/cta-observatory/pyirf}{https://github.com/cta-observatory/pyirf}}~\citep{Noethe2021, pyirf}, a prototype for the generation of IRFs and sensitivities for CTA. The CTA analysis workflow with conventional methods can be found in these proceedings in Ref.~\citep{Noethe2021}.

\paragraph{Full-event reconstruction with CTLearn}

\begin{figure}[h]
    \centering
    \includegraphics[width=0.8\textwidth]{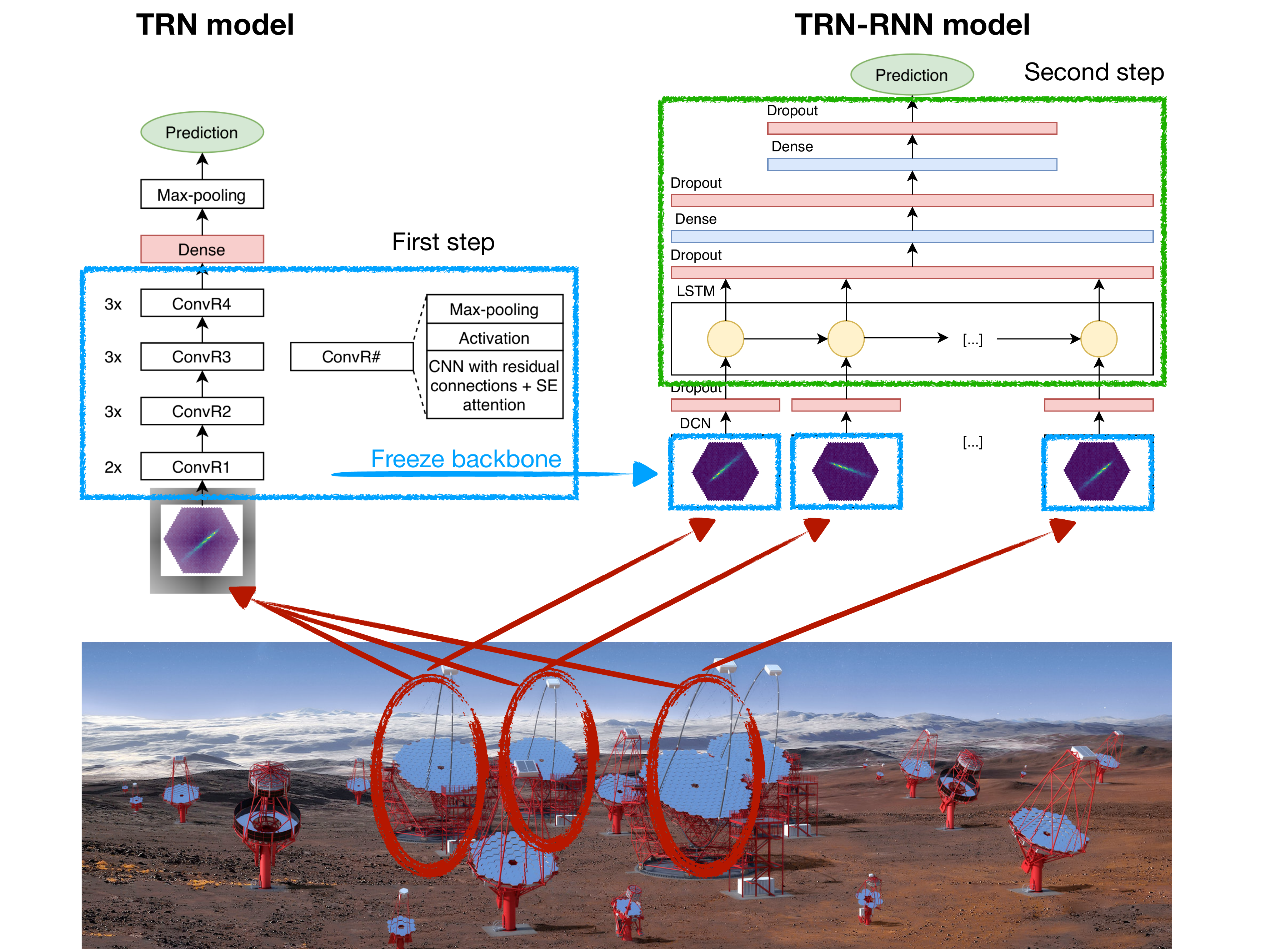}
    \caption{Diagram depicting the main layers of the {\em TRN} (left) and the {\em TRN-RNN} model (right).}
    \label{fig:ctlearn-models}
\end{figure}

The high-level, open-source \texttt{CTLearn} package provides a framework for training deep learning models for IACT full-event reconstruction using \texttt{TensorFlow}. This work focuses on the thin ResNet ({\em TRN}) model \citep{2019arXiv190210107X} (see Fig.~\ref{fig:ctlearn-models}, left model), a deep DCN-based architecture for monoscopic full-event reconstruction with residual connections (meaning that the original input is added to the output at each stage, demonstrated to improve performance) \citep{2015arXiv151203385H}. A dual (squeeze-and-excitation) attention mechanism \citep{2017arXiv170901507H} is deployed in each of the residual blocks. Either particle classification or regression (energy or arrival direction reconstruction) is performed with a selectable fully-connected head (FCH), a traditional multi-layer perceptron neural network (MLP), after the deep backbone, consisting of several stacked residual blocks.

The stereoscopic full-event reconstruction is performed with the {\em TRN-RNN} model (see Fig.~\ref{fig:ctlearn-models}, right model), which consists of multiple thin ResNet blocks connected via a recurrent neural network (RNN)~\citep{{Goodfellow-et-al-2016}}. In particular, the RNN - implemented in \texttt{CTLearn} - is a dynamic long short-term memory network (LSTM)~\citep{{Goodfellow-et-al-2016}}, which adjusts its size according to the number of triggered telescopes for each event. The images of triggered telescopes are sorted a priori by the total amount of integrated charge in the camera. To train this large architecture, and to overcome computing limitations, transfer learning~\citep{{Goodfellow-et-al-2016}} is utilized: the DCN backbone of the {\em TRN} model is trained beforehand, its parameters are set to be untrainable weights and transferred into the {\em TRN-RNN} model.

\section{Dataset}
\label{sec:dataset}

The analysis is carried out with the CTA South (zenith angle of $ 20^{\circ} $, North pointing) reference dataset, processed with \texttt{ctapipe}.\footnote{The first stage of the analysis was performed on the EGI (\href{www.egi.eu}{www.egi.eu}).} A detailed description of the simulation production, together with the telescope layout and performance study of CTA can be found elsewhere in Ref.~\citep{Gueta2021}. For the deep learning training process of the particle classification,  diffuse gamma-ray and proton-initiated events, simulated within a cone of 10$^{\circ}$ radius (covering the whole field of view of the instrument) are considered, in a balanced way so both populations contribute equally to the statistics of the datasets. 80\% of the data are used for training (from which 5\% are reserved for validation of the learning process) and 20\% for testing. The regression models (energy or arrival direction reconstruction) are trained with the whole training set of diffuse gamma rays. The performance of the deep learning models is evaluated on simulated protons ($ \sim 7\mathrm{e}^{9} $), electrons ($ \sim 2\mathrm{e}^{9} $) and point-source gamma rays ($ \sim 2\mathrm{e}^{9} $) with $ 0.4^{\circ} $ offset with respect to the telescope pointing.

\paragraph{Array layout}
The array layout \texttt{M5C5}, consists of 13 medium-size telescopes (MSTs) and 40 small-size telescopes (SSTs), and four additional large-size telescopes (LSTs) are considered in this work. The \texttt{M6C5} array layout with 14 MSTs, 40 SSTs and no LSTs is depicted in lower left panel of Fig. 1 in Ref.~\citep{Gueta2021}. For the monoscopic full-event reconstruction, all images from the corresponding telescope type, regardless of the particular telescope, are included at training stage. However, the single telescope performances are evaluated with one particular telescope per telescope type. 

\paragraph{Data selection (quality cuts)}
In order to compare to the conventional IACT analysis methods, two different data selection cuts are performed.\footnote{Please note that DCNs are fed with all the information contained in the event images and therefore no default quality cuts have to be applied.} The {\em TRN} model are trained and tested with a modest cut, adapted from Ref.~\citep{Grespan2021}, where faint images (Hillas intensity less than 50 photoelectrons) and images close to the camera edge (leakage2 parameter more than 0.2) are discarded. For the stereoscopic reconstruction, faint and truncated images are kept, but a multiplicity cut of four or more triggered telescopes is applied.

\section{Results}
\label{sec:results}

The \emph{TRN} and \emph{TRN-RNN} models successfully learn to perform monoscopic and stereoscopic full-event reconstruction, respectively, for all sizes of CTA telescopes. The standard IACT IRFs and sensitivities are depicted in Fig.~\ref{fig:trn-results} and~\ref{fig:trn-rnn-results}. The angular resolution is defined as the angle containing 68\% of the reconstructed gamma-ray events relative to the simulated point source gamma-ray direction.  This is calculated in each logarithmic energy bin. The energy resolution in each energy bin is calculated with 68\% of containment of $ (E_{\mathrm{reco}} - E_{\mathrm{true}})/E_{\mathrm{true}} $. The effective collection area, which is proportional to the gamma-ray efficiency of detection, is computed as a function of the simulated energy. Only events entering the calculation of the sensitivity curve are considered for the effective collection area and the resolution curves. The differential sensitivity calculation requires a minimal significance of more than 5 $ \sigma $, at least ten detected gamma rays and a minimal excess over background ratio of 0.05 for a observation of 50 hours. For IACTs, the receiver operating characteristic (ROC) curve visualizes the diagnostic ability of the gamma/hadron separation as its gammaness threshold is varied. The area under the ROC curve (AUC) is a measurement of the quality of the background rejection.

\subsection{Single-telescope event reconstruction ({\em TRN} model)}

\begin{figure}[h]
    \centering
    \includegraphics[width=0.725\textwidth]{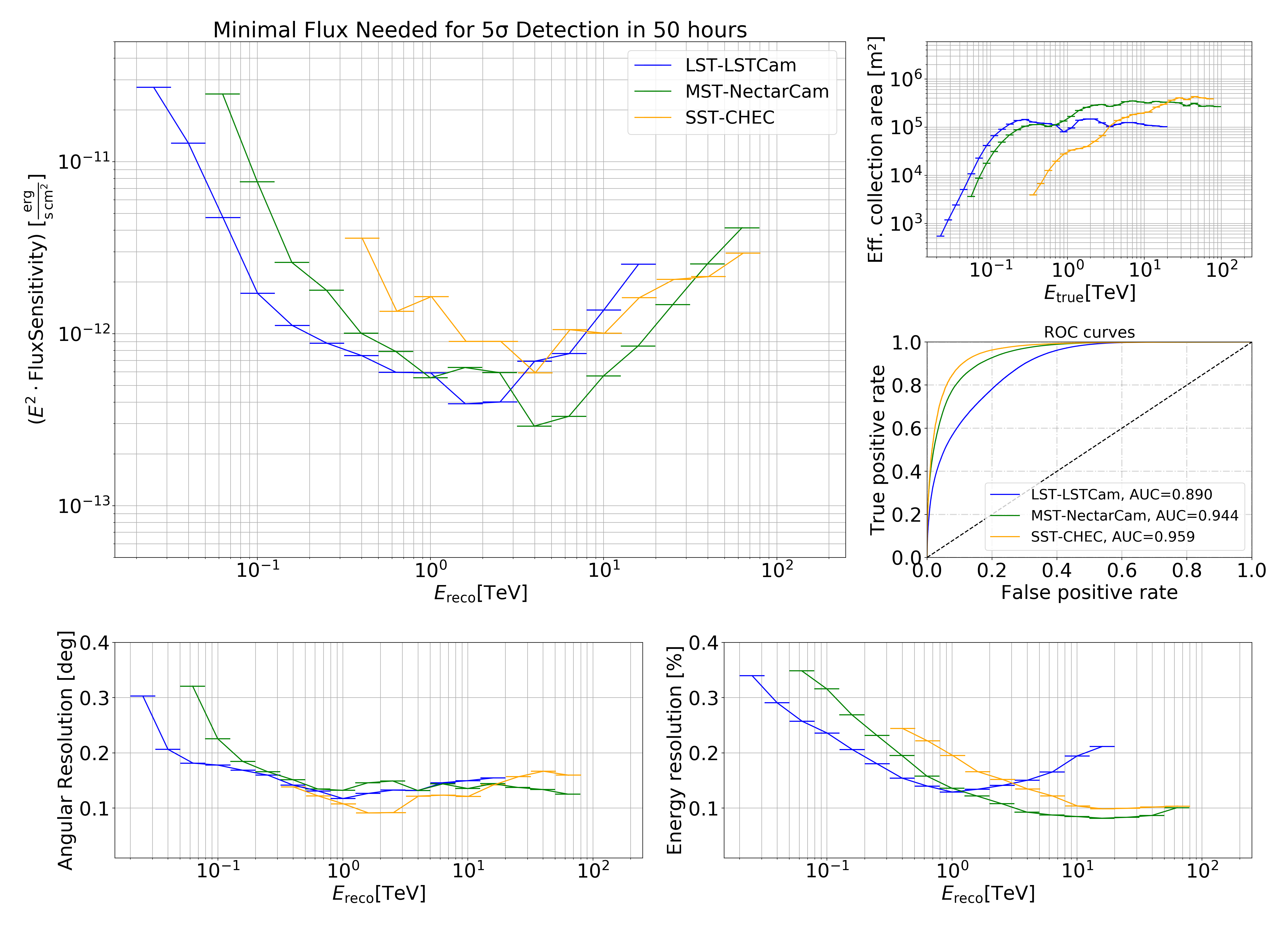}
    \caption{The single telescope IRFs and sensitivities, as defined in Sec.~\ref{sec:results}, obtained with the {\em TRN} model for the LST (blue), MST (green), and SST (orange).}
    \label{fig:trn-results}
\end{figure}

The \emph{TRN} model is trained on $ \sim $ 200k batches of 64 images for each telescope type, validating periodically. As expected (see Fig.~\ref{fig:trn-results}), the LST is the most sensitive telescope type at the lowest energies; the sensitivity of an MST is best where this telescope type will be responsible for the full-array sensitivity; the SST is providing competitive sensitivity at multi-TeV energies. In the mono telescope analysis, SSTs are competitive with MSTs only at the highest energies. The LST, MST, and SST provide excellent energy resolutions of $ \sim $ 13\%, 9\%, and 10\% at their best, angular resolutions of $ \sim 0.12^{\circ}$, $ 0.13^{\circ}$ and $ 0.1^{\circ}$ at their best, and an AUC of 0.89, 0.944, and 0.959, in their entire energy ranges, respectively.

\subsection{Reconstruction of stereoscopic events ({\em TRN-RNN} model)}

\begin{figure}[h]
    \centering
    \includegraphics[width=0.725\textwidth]{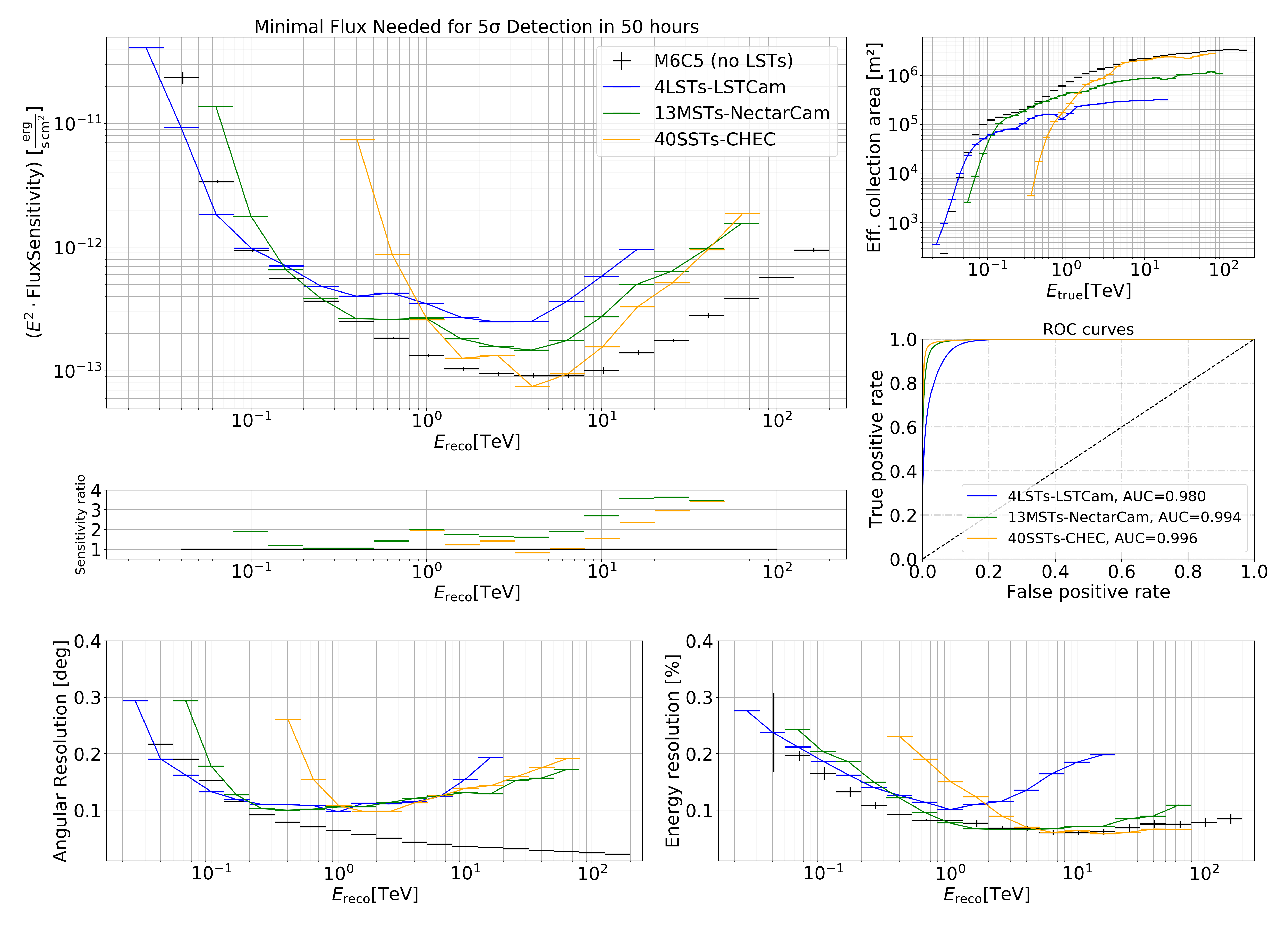}
    \caption{The multi telescopes IRFs and sensitivities, as defined in Sec.~\ref{sec:results}, obtained with the {\em TRN-RNN} model for 4~LSTs (blue), 13~MSTs (green), and 40~SSTs (orange). The black curves depict the IRFs and sensitivities obtained with the conventional analysis of the array layout \texttt{M6C5} with 14~MSTs, 40~SSTs, and no LSTs taken from Ref.~\citep{Gueta2021}.}
    \label{fig:trn-rnn-results}
\end{figure}

As discussed in Sec.~\ref{sec:workflow}, the backbone of the \emph{TRN} model is transferred into the \emph{TRN-RNN} model. Therefore, only $ \sim $ 100k batches of 16 images for each telescope type are needed to train the RNN and the FCH blocks of the model. The learning is also validated periodically. The particle classification, performed by the \emph{TRN-RNN} model (see Fig.~\ref{fig:trn-rnn-results}), works well, with an AUC score of 0.98, 0.994, and 0.996 for the subarrays of 4~LSTs-LSTCam, 13~MSTs-NectarCam, and 40~SSTs-CHEC, respectively. The three subarrays reach promising top values for the the energy resolution of $ \sim $ 10\%, 7\%, and 6\% for LSTs, MSTs and SSTs, respectively. The IRFs produced with our models are cut off above 80 TeV, likely because they fail the requirement of 10 gamma rays in each energy bin. This should be solved once all telescope types would be combined in the reconstruction.

The \emph{TRN-RNN} model performs poorly on the reconstruction of the arrival direction. The angular resolution for the highest energies differs from the conventional analysis significantly, which translates to the sensitivity curves, causing a deficit of performance especially at energies above 10 TeV. Achieving just a small improvement for the angular resolution by adding more telescopes concludes that this version of the \emph{TRN-RNN} model needs further adjustments to be fully capable of learning stereoscopic features relevant for the arrival direction reconstruction. Including further information like the telescope position may help the network to better reconstruct the arrival direction. Future studies are planned to improve the angular resolution with DCN-based models especially in stereo mode. A fair comparison to the conventional analysis with the Eventdisplay software package (see black curves in Fig.~\ref{fig:trn-rnn-results}) is not feasible at this stage of the development, because no LSTs are considered in the conventional analysis, and it is not limited to a per-telescope-type analysis \citep{Gernot2017}.

\section{Conclusion and Outlook}
\label{sec:conclusion}

This contribution shows for the first time that DCN-based full-event reconstruction works for all sizes of CTA telescopes, in both single-telescope and stereo modes. The performance of the \emph{TRN} and the \emph{TRN-RNN} models for the particle classification and the energy estimation is promising. Tackling the arrival direction reconstruction task via DCNs requires additional modifications and improvements to the existing stereoscopic deep learning models to suit the requirements of CTA.

Future developments of CTLearn will include the combination of different telescope types to evaluate the full-array performance of CTA North and South with deep learning models. The results of each telescope type are obtained with the same set of non-optimized hyperparameters. Hyperparameter optimization will be explored in future. Multitask learning experiments (see Ref.~\citep{Vuillaume2021}), where one single model performs the IACT specific tasks (particle classification, energy and arrival direction estimation), as well as the application of DCNs model on observational data, are also planned. Further validation of DCN-based full-event reconstruction under various circumstances (i.e. off-axis performance, divergent pointing, different zenith angles and night sky backgrounds, etc.) is very important and will be considered in future works.

\section{Acknowledgments}
\label{sec:acknowledgments}
This work was conducted in the context of the CTA Analysis and 
Simulations Working Group. We gratefully acknowledge financial support from the 
agencies and organizations listed in this \href{http://www.cta-observatory.org/consortium_acknowledgments}{link}.
TM acknowledges support from PID2019-104114RB-C32. DN and JLC acknowledges partial support from The European Science Cluster of Astronomy 
\& Particle Physics ESFRI Research Infrastructures funded by the European Union’s 
Horizon 2020 research and innovation program under Grant Agreement no. 824064. This work was performed using resources provided by the Cambridge Service for Data Driven Discovery (CSD3) operated by the University of Cambridge Research Computing Service (www.csd3.cam.ac.uk), provided by Dell EMC and Intel using Tier-2 funding from the Engineering and Physical Sciences Research Council (capital grant EP/P020259/1), and DiRAC funding from the Science and Technology Facilities Council (STFC, www.dirac.ac.uk).This work used IRIS computing resources funded by the STFC. SS acknowledges an STFC PhD studentship. We acknowledge the support of NVIDIA Corporation with the donation of a Titan X Pascal GPU used for part of this research.
\\
\\
This paper has gone through internal review by the CTA Consortium.

%% Full authors list (ONLY FOR COLLABORATIONS)
\clearpage
\section*{The Cherenkov Telescope Array Consortium July 2021 Authors}

%\noindent \textbf{Note comment afterwards:} Collaborations have the possibility to provide an authors list in xml format which will be used while generating the DOI entries making the full authors list searchable in databases like Inspire HEP. For instructions please go to icrc2021.desy.de/proceedings or contact us under icrc2021proc@desy.de.\\
%
%\scriptsize
%\noindent
%first.author$^1$, 
%second.author$^2$, 
%third.author$^3$ % .... more names
%and 
%last.author$^{n}$ \\
%
%\noindent
%$^1$first.affiliation.
%$^2$second.affiliation. % .... more affiliation
%$^{m}$last.affiliation.

\scriptsize
\noindent
H.~Abdalla\textsuperscript{1}, H.~Abe\textsuperscript{2},
S.~Abe\textsuperscript{2}, A.~Abusleme\textsuperscript{3},
F.~Acero\textsuperscript{4}, A.~Acharyya\textsuperscript{5}, V.~Acín
Portella\textsuperscript{6}, K.~Ackley\textsuperscript{7},
R.~Adam\textsuperscript{8}, C.~Adams\textsuperscript{9}
% ...

1 : Centre for Space Research, North-West University, Potchefstroom, 2520, South Africa

2 : Institute for Cosmic Ray Research, University of Tokyo, 5-1-5, Kashiwa-no-ha, Kashiwa, Chiba 277-8582, Japan

3 : Pontificia Universidad Católica de Chile, Av. Libertador Bernardo O'Higgins 340, Santiago, Chile

4 : AIM, CEA, CNRS, Université Paris-Saclay, Université Paris Diderot, Sorbonne Paris Cité, CEA Paris-Saclay, IRFU/DAp, Bat 709, Orme des Merisiers, 91191 Gif-sur-Yvette, France

5 : Centre for Advanced Instrumentation, Dept. of Physics, Durham University, South Road, Durham DH1 3LE, United Kingdom

6 : Port d'Informació Científica, Edifici D, Carrer de l'Albareda, 08193 Bellaterrra (Cerdanyola del Vallès), Spain

7 : School of Physics and Astronomy, Monash University, Melbourne, Victoria 3800, Australia

8 : Laboratoire Leprince-Ringuet, École Polytechnique (UMR 7638, CNRS/IN2P3, Institut Polytechnique de Paris), 91128 Palaiseau, France

9 : Department of Physics, Columbia University, 538 West 120th Street, New York, NY 10027, USA

% ...


\begin{thebibliography}{99}

\scriptsize
%\tiny

\bibitem{Acharyya:2019nwy} Acharyya, A. {\it et al.}, {\em Monte Carlo studies for the optimisation of the Cherenkov Telescope Array layout}, Astropart.\ Phys.\  {\bf 111}, 35 (2019), \href{https://doi.org/10.1016/j.astropartphys.2019.04.001}{doi:10.1016/j.astropartphys.2019.04.001}


\bibitem{1985ICRC....3..445H}
{Hillas},A.~M., {\it {Cerenkov light images of EAS produced by primary gamma}},
   {\em International Cosmic Ray Conference} {\bf 3} (1985)

\bibitem{2008NIMPA.588..424A}
{Albert}, J., {\it et al.}, {\em Implementation of the Random Forest method for the
  Imaging Atmospheric Cherenkov Telescope MAGIC}, Nuclear Instruments
  and Methods in Physics Research A {\bf 588} 424--432 (2008),
  [\href{http://arxiv.org/abs/0709.3719}{{\tt arXiv:0709.3719}}]

\bibitem{2009APh....31..383O} Ohm, S., van Eldik, C., Egberts, K., {\em {\ensuremath{\gamma}}/hadron separation in very-high-energy {\ensuremath{\gamma}}-ray astronomy using a multivariate analysis method}, Astroparticle Physics 31, 383 (2009)

\bibitem{2011APh....34..858B}
{Becherini}, Y., {Djannati-Ata{\"i}}, A., {Marandon}, V., {Punch}, M., and
  {Pita}, S., {\it {A new analysis strategy for detection of faint
  {$\gamma$}-ray sources with Imaging Atmospheric Cherenkov Telescopes}},  {\em
  Astroparticle Physics} {\bf 34} 858--870 (2011),
  [\href{http://arxiv.org/abs/1104.5359}{{\tt arXiv:1104.5359}}]
  
\bibitem{krause2017improved}
Krause, M., Pueschel, E., and Maier, G., {\it Improved $\gamma$/hadron separation
  for the detection of faint $\gamma$-ray sources using boosted decision
  trees},  {\em Astroparticle Physics} {\bf 89} 1--9 (2017)

\bibitem{Goodfellow-et-al-2016}
Goodfellow, I.,  Bengio, Y. and Courville, A., {\em Deep Learning}, MIT Press (2016), \href{http://www.deeplearningbook.org}{www.deeplearningbook.org}

\bibitem{2017arXiv170905889N}  Nieto, D., Brill, A., Kim, B., Humensky, T.~B., {\em et al.}, {\it Exploring deep learning as an event classification method for the Cherenkov Telescope Array}, 35th International Cosmic Ray Conference (2017), \href{https://doi.org/10.22323/1.301.0809}{doi:10.22323/1.301.0809}

\bibitem{veritas_cnn}
{Feng} Q, and {Lin}, T.~T.~Y., {\it {The analysis of VERITAS muon images using
  convolutional neural networks}},  {\em ArXiv e-prints} (2016)
  [\href{http://arxiv.org/abs/1611.09832}{{\tt arXiv:1611.09832}}]
  
\bibitem{2018arXiv181000592M} Mangano, S., {\it et al.}, {\em Extracting gamma-ray information from images with convolutional neural network methods on simulated Cherenkov Telescope Array data}, ANNPR 2018, LNAI 11081 243-254 (2018),  \href{https://dx.doi.org/10.1007/978-3-319-99978-4}{doi:10.1007/978-3-319-99978-4}

\bibitem{2019APh...105...44S} Shilon, I., {\it et al.}, {\em Application of deep learning methods to analysis of imaging atmospheric Cherenkov telescopes data}, Astroparticle Physics 105, 44 (2019)

\bibitem{2019-icrc-hex} Nieto, D., {\it et al.}, {\em Studying deep convolutional neural networks with hexagonal lattices for imaging atmospheric Cherenkov telescope event reconstruction}, {\em 36th International Cosmic Ray Conference} (2019)

\bibitem{2021spencer} Spencer, S., {\it et al.}, {\em Deep learning with photosensor timing information as a background rejection method for the Cherenkov Telescope Array}, Astroparticle Physics {\em ArXiv e-prints} (2017)
  [\href{http://arxiv.org/abs/2103.06054}{{\tt arXiv:2103.06054}}]

\bibitem{Grespan2021} Grespan, P., {\it et al.}, {\em Deep-learning-driven event reconstruction applied to simulated data from a single Large-Sized Telescope of CTA}, {\em 37th International Cosmic Ray Conference} (2021)

\bibitem{Vuillaume2021} Vuillaume, T., {\it et al.}, {\em Analysis of the Cherenkov Telescope Array first Large Size Telescope real data using convolutional neural networks}, {\em 37th International Cosmic Ray Conference} (2021)


\bibitem{Noethe2021} Noethe, M., {\it et al.}, {\em Prototype Open Event Reconstruction Pipeline for CTA}, {\em 37th International Cosmic Ray Conference} (2021)

\bibitem{ctapipe} {\em cta-observatory/ctapipe: v0.10.5}, Zenodo (2021) \href{https://doi.org/10.5281/zenodo.4581045}{10.5281/zenodo.4581045}, \href{https://github.com/cta-observatory/ctapipe}{https://github.com/cta-observatory/ctapipe}.

\bibitem{dl1dh} {\em DL1 Data Handler: HDF5 format IACT data writer, reader and processor}, Zenodo (2019) \href{https://doi.org/10.5281/zenodo.3336560}{doi:10.5281/zenodo.3336560}, \href{https://github.com/cta-observatory/dl1-data-handler}{https://github.com/cta-observatory/dl1-data-handler}.

\bibitem{ctlearn} {\em CTLearn: Deep Learning for IACT Event Reconstruction}, Zenodo (2019), \href{https://doi.org/10.5281/zenodo.3342952}{doi:10.5281/zenodo.3342952}, \href{https://github.com/ctlearn-project/ctlearn}{github.com/ctlearn-project/ctlearn}

\bibitem{2019-icrc-ctlearn} Nieto, D., {\it et al.}, {\em CTLearn: Deep Learning for Gamma-ray Astronomy}, {\em 36th International Cosmic Ray Conference} (2019)

\bibitem{pyirf} {\em cta-observatory/pyirf: v0.5.0}, Zenodo (2021) \href{https://doi.org/10.5281/zenodo.4748994}{doi:10.5281/zenodo.4748994}, \href{https://github.com/cta-observatory/pyirf}{https://github.com/cta-observatory/pyirf}.


\bibitem{2019arXiv190210107X} Xie, W., {\it et al.}, {\em Utterance-level Aggregation For Speaker Recognition In The Wild}, Electrical Engineering and Systems Science - Audio and Speech Processing, Computer Science - Machine Learning, Computer Science - Multimedia, Computer Science - Sound {\em ArXiv e-prints} (2017)
  [\href{http://arxiv.org/abs/1902.10107}{{\tt arXiv:1902.10107}}]

\bibitem{2015arXiv151203385H} He, K., {\it et al.}, {\em Deep Residual Learning for Image Recognition}, Computer Science - Computer Vision and Pattern Recognition {\em ArXiv e-prints} (2017)
  [\href{http://arxiv.org/abs/1512.03385}{{\tt arXiv:1512.03385}}]

\bibitem{2017arXiv170901507H} Hu, J., {\it et al.}, {\em Squeeze-and-Excitation Networks}, Computer Science - Computer Vision and Pattern Recognition {\em ArXiv e-prints} (2017)
  [\href{http://arxiv.org/abs/1709.01507}{{\tt arXiv:1709.01507}}]
 
\bibitem{monoCTLearn} Nieto, D., {\it et al.}, {\em Reconstruction of IACT events using deep learning techniques with CTLearn}, {\em 30th Astronomical Data Analysis Software and Systems (ADASS) conference} (2020)

\bibitem{Gueta2021} Gueta, O., {\it et al.}, {\em The Cherenkov Telescope Array: layout, design and performance}, {\em 37th International Cosmic Ray Conference} (2021)

\bibitem{Gernot2017} Maier, G., {\it et al.}, {\em Eventdisplay: An Analysis and ReconstructionPackage for Ground-based Gamma-ray Astronomy}, {\em 35th International Cosmic Ray Conference} (2017)

\end{thebibliography}
\end{document}